\documentclass{aa}
\usepackage{graphicx,psfig,eucal}
\begin{document}
   \title{The black hole mass of low redshift radiogalaxies}


   \author{D. Bettoni \inst{1}, R. Falomo \inst{1}, G. Fasano  \inst{1},
F. Govoni \inst{2}}

   \offprints{D. Bettoni}

   \institute{$^{(1)}$INAF - Osservatorio Astronomico di Padova 
 vicolo Osservatorio 5 Padova\\
     \email{bettoni@pd.astro.it, falomo@pd.astro.it, 
      fasano@pd.astro.it} \\
$^{(2)}$Istituto di Radioastronomia di Bologna and Dipartimento di Astronomia,
Universit\'a di Bologna
\email{fgovoni@ira.bo.cnr.it} \\
             }

   \date{Received; accepted }
\authorrunning{Bettoni et al.}

   \abstract{We make use of two empirical relations between the black
   hole mass and the global properties (bulge luminosity and stellar
   velocity dispersion) of nearby elliptical galaxies, to infer the
   mass of the central black hole ($\CMcal{M}_{BH}$) in low redshift
   radiogalaxies.  Using the most recent determinations of black hole
   masses for inactive early type galaxies we show that the bulge
   luminosity and the central velocity dispersion are almost equally
   correlated (similar scatter) with the central black-hole
   mass. Applying these relations to two large and homogeneous
   datasets of radiogalaxies we find that they host black-holes whose
   mass ranges between $\sim 5\times10^7$ to $\sim 6\times10^9\CMcal
   {M}_{\odot}$ (average $<Log\CMcal{M}_{BH}>
   \sim$8.9). $\CMcal{M}_{BH}$ is found to be proportional to the mass
   of the bulge ($\CMcal{M}_{bulge}$).  The distribution of the ratio
   $\CMcal{M}_{BH}$/$\CMcal{M}_{bulge}$ has a mean value of
   8$\times10^{-4}$ and shows a scatter that is consistent with that
   expected from the associated errors. At variance with previous
   claims no significant correlation is instead found between
   $\CMcal{M}_{BH}$ (or $\CMcal{M}_{bulge}$) and the radio power at 5
   GHz.   }

   \maketitle
%

\section{Introduction}

There is a large consensus about the existence of super massive black
holes (SBHs) at the center of nearby inactive galaxies as well as in
the nuclei of active galaxies and quasars (see e.g. for a recent
review \cite{ferrarese}).  A large body of data, in particular based on high
resolution HST observations, is now available (see e.g. 
 \cite{kormendy2}) to support the presence of such
massive black holes (BH) using different techniques.

It is believed that SBHs play an important role in the formation
and evolution of massive galaxies and also to be a key component for the
development of the nuclear activity. However in spite of this apparently
ubiquitous presence of SBHs in galaxies, our understanding on how the
galaxies and their central BHs are linked in the process of formation
of the observed structures is still poor (see \cite{silk};
\cite{haehnelt}; \cite{adams}).

The most important result obtained from the measured BH masses in
nearby galaxies is the existence of a significant correlation between the
black hole mass ($\CMcal{M}_{BH}$) and the mass of the bulge component
($\CMcal{M}_{bulge}$) of the host galaxy ($\CMcal{M}_{bulge}$). From
the observational point of view this correlation is translated into
relationships between $\CMcal{M}_{BH}$ and bulge luminosity
$L_{bulge}$ (\cite{magorrian}, \cite{kormendy2}) and between
$\CMcal{M}_{BH}$ and the stellar velocity dispersion $\sigma$ (\cite{FM};
\cite{gebhardt} ).

Although based on a small number ($\sim$30) of nearby galaxies for
which direct dynamical measurements of $\CMcal M_{BH}$ have been
secured, and in spite of their scatter [$\sim$0.4 in $Log
(\CMcal{M}_{BH})$], these empirical relationships offer a new tool for
evaluating $\CMcal{M}_{BH}$ in various types of AGN, provided that
bulge luminosities and/or velocity dispersion be available (see also
\cite{MD}, \cite{FCT}).

In this paper we make use of such relationships to investigate
the BH mass distribution of two large and homogeneous datasets of low
redshift radiogalaxies (RG) for which we have previously studied the
morphological, structural, photometrical and kinematical properties
(\cite{FFS}; ~\cite{fede1},b; ~\cite{bettoni}).  The
derived BH masses of radiogalaxies are then used to investigate the
connections between $\CMcal{M}_{BH}$, the mass of the galaxy and the
radio power.

To this aim we first describe our samples (Section 2) and revisit the
relations $\CMcal{M}_{BH}$-$L_{bulge}$ and $\CMcal{M}_{BH}$-$\sigma$
for nearby early-type galaxies (Section 3). Then we use these
relationships to evaluate $\CMcal{M}_{BH}$ of radio galaxies (Section
4) and to study the connections between $\CMcal{M}_{BH}$ and the mass
of the bulge component of the host galaxy and between $\CMcal{M}_{BH}$
and the radio luminosity. A summary of the main conclusions of this
study is reported in Section 5. In our analysis we assume H$_{0}$=50
Km s$^{-1}$ Mpc$^{-1}$ and $\Omega_{0}$=0.


\section{The Samples}

\subsection{The sample of inactive nearby ellipticals}

\begin{table*}
      \caption[]{Properties of the nearby inactive galaxy sample}
         \label{ngc_bh}
\begin{tabular}{llrccccccc}
\hline
{\rm Object}& Type & D & $B_t$ & $B-R$ &$A_B$ & M$_B$ & $\sigma$ & Log$R_e$ & Log($\CMcal{M}_{BH}$) \\
& & Mpc & (km$s^{-1}$) &  & & & (km$s^{-1}$)& (kpc) & $\CMcal {M}_{\odot}$ \\
(1)&(2) & (3) & (4) &(5) &(6) &(7) &(8) & (9) &(10)  \\
\hline
N221/M32 & cE2 &0.81 &  8.52&  1.53 &  0.55&  -16.02 &  76 & -0.82 &  6.40  \\
  N821 &  E6 & 24.10 & 11.49&  1.68 &  0.55&  -20.42 & 196 &  0.72 &  7.57  \\
 N2778 &   E & 22.90 & 13.22&  1.64 &  0.17&  -18.58 & 171 &  0.26 &  7.15  \\
 N3377 & E5+ & 11.20 & 11.07&  1.37 &  0.21&  -19.18 & 131 &  0.26 &  8.00  \\
 N3379 &  E1 & 10.60 & 10.32&  1.57 &  0.21&  -19.81 & 210 &  0.26 &  8.00  \\
 N3608 &  E2 & 23.00 & 11.64&  1.51 &  0.13&  -20.17 & 206 &  0.59 &  8.28  \\
 N4261 &  E2 & 31.60 & 11.27&  1.61 &  0.13&  -21.23 & 290 &  0.77 &  8.72  \\
 N4291 &   E & 26.20 & 12.37&  1.60 &  0.21&  -19.72 & 269 &  0.27 &  8.49  \\
 N4374 &  E1 & 18.40 & 10.09&  1.61 &  0.21&  -21.23 & 286 &  0.69 &  8.63  \\
 N4473 &  E5 & 15.70 & 11.12&  1.56 &  0.21&  -19.86 & 188 &  0.28 &  8.04  \\
 N4486 &  E1 & 16.70 &  9.40&  1.63 &  0.21&  -21.71 & 345 &  0.92 &  9.48  \\
 N4564 &   E & 15.00 & 11.94&  1.57 &  0.17&  -18.94 & 153 &  0.20 &  7.75  \\
 N4649 &  E2 & 16.80 &  9.72&  1.64 &  0.17&  -21.41 & 331 &  0.78 &  9.30  \\
 N4697 &  E6 & 11.70 &  9.99&  1.54 &  0.17&  -20.35 & 163 &  0.63 &  8.23  \\
 N4742 &  E4 & 15.50 & 11.92&  1.33 &  0.29&  -19.03 &  93 & -0.05 &  7.15  \\
 N5128 &$S0_{pec}$ &4.20 & 7.30& 1.60 &  0.50&  -20.82 & 145 &  0.82$^{(a)}$ & 8.38  \\
 N5845 &  E* & 25.90 & 13.15&  1.65 &  0.34&  -18.92 & 275 & -0.29 &  8.38  \\
 N6251 &   E & 106.00& 13.18&  1.60 &  0.42&  -21.95 & 297 &  1.31$^{(b)}$ &  8.72  \\
 N7052 &   E & 58.70 & 12.41&  1.60 &  0.80&  -21.43 & 261 &  0.82${(c)}$ &  8.52  \\
 I1459 &  E3 & 29.20 & 10.88&  1.59 &  0.07&  -21.45 & 312 &  0.74 &  9.40  \\
\hline
\end{tabular}
\begin{list}{}{}
\item $^{(a)}$ from Dufour et al. (1979), $^{(b)}$ from Owen \& Laing (1989), 
$^{(c)}$ from Gonzalez-Serrano et al. (1993). 
\end{list}
  \end{table*}

In order to investigate the relations $\CMcal{M}_{BH}$-$L_{bulge}$ 
and $\CMcal{M}_{BH}$-$\sigma$ of E-type galaxies we have considered a
sample of 20 objects of E-type morphology (excluding lenticulars) in
the Kormendy \& Gebhardt (2001) galaxy list with measured BH
masses. The Milky Way was therefore not considered.  In Table 1 we
report the relevant data for this sample: columns 1 and 2 give the
name and the morphological type from RC3; columns 3 the distance,
derived from Surface Brightness Fluctuations (SBF, \cite{tonry});
columns 4, 5, 6, 7 report the adopted apparent total B magnitude
corrected for extinction, the $B-R$ color, the galactic extinction,
derived from the Bell Laboratories Survey of neutral Hydrogen
(\cite{stark}) and the absolute B magnitude respectively. The $B-$band
bulge magnitudes were taken from Faber et al. (1997) for all the 20
objects but for NGC~5128, NGC~6251 and NGC~7052 we adopted the
magnitudes given in RC3 (\cite{devoc}). Note that for this sample the
bulge magnitude is coincident with the total magnitude of the galaxy
since the galaxy luminosity profile is always well represented by an
$r^{1/4}$ law. The adopted values for the velocity dispersion (from
\cite{FM}), effective radius $R_e$ (from Faber et al. 1989 and using
the distance in column 3) and $\CMcal{M}_{BH}$ are reported in columns
8, 9 and 10.  The latter values are taken from Tremaine et al. (2002)
that give the most recent revision of $\CMcal{M}_{BH}$ for this data
set.

\subsection{The radio-galaxies sample}

\begin{figure*}
\includegraphics[width=14cm]{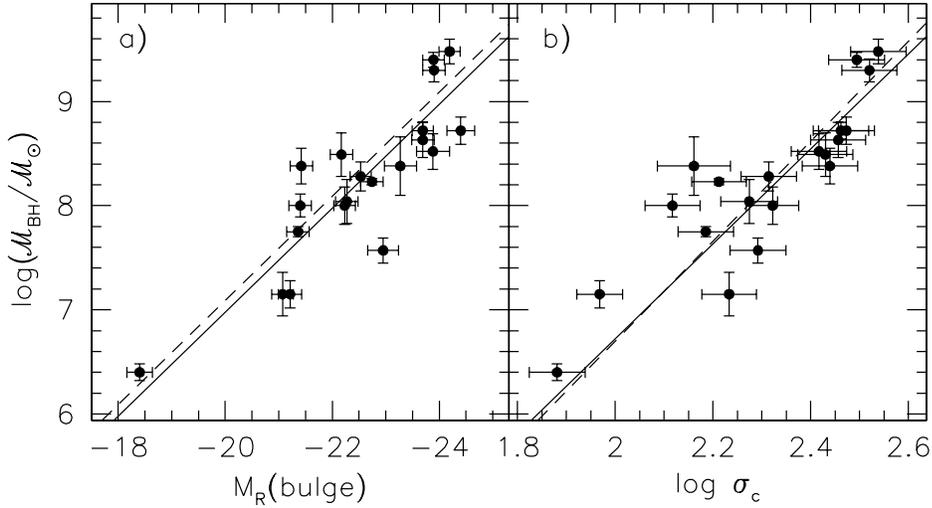}
\vspace{-2.5cm}
\caption{{\it a}) The $\CMcal{M}_{BH}$ vs $M_{Bulge}$ relation for the
20 nearby inactive ellipticals (see Table~\ref{ngc_bh}); {\it b}) Same
as {\it a}) but for the $\CMcal{M}_{BH}$ vs Log$(\sigma)$
relation. The solid lines refer to our linear best fits. The dashed
lines represent the relation of McLure and Dunlop (2002) scaled to our
cosmology (panel {\it a}) and that of Ferrarese \& Merritt
(2000)(panel {\it b}).}
\label{MBH-MB-sigma}
\end{figure*}

\begin{table*}
      \caption[]{The Sample A of radio galaxies}
         \label{rg_bh}
\begin{tabular}{llcccccccc}
\hline
{\rm Object}&  z & M$_R$ & Log($\sigma$) &  Log($L_{48}$) &  Log($L_{48}$) & Ref.$^{a}$ & Log$\CMcal{M}_{BH}$ & Log$\CMcal{M}_{BH}$& Log$\CMcal {M}_{bulge}$ \\
& & host &  & total   & core  & &  from $\sigma$ & from M(bulge)  & \\
& & & km$s^{-1}$ &  (W/Hz) &  (W/Hz) & & $\CMcal{M}_{\odot}$ & $\CMcal{M}_{\odot}$  & $\CMcal{M}_{\odot}$ \\
(1)&(2) & (3) & (4) &(5) &(6) &(7) &(8) & (9) &(10) \\
\hline 			      
 0055-016 & 0.045 & -24.08 &2.48 &  25.30 &  23.93  & WP&  9.00 & 8.99 & 12.10    \\
 0131-367 & 0.030 & -24.10 &2.40 &  25.21 &  23.18  & WP&  8.64 & 9.01 & 11.92    \\
 0257-398 & 0.066 & -23.73 &2.34 &  24.94 & $<$22.9 & EK&  8.79 & 8.80 & 11.75    \\
 0312-343 & 0.067 & -24.02 &2.41 &  24.66 &  23.71  & EK&  8.68 & 8.95 & 12.18    \\
 0325+023 & 0.030 & -23.41 &2.34 &  24.90 &  23.80  & WP&  8.37 & 8.66 & 11.93    \\
 0449-175 & 0.031 & -23.97 &2.20 &  24.12 &  22.63  & EK&  8.04 & 8.94 & 11.73    \\
 0546-329 & 0.037 & -24.49 &2.59 &  24.13 &  23.29  & EK&  9.49 & 9.20 & 12.46    \\
 0548-317 & 0.034 & -23.35 &2.09 &  24.35 &$<$22.49 & EK&  9.31 & 8.63 & 12.12    \\
 0718-340 & 0.029 & -24.20 &2.52 &  24.55 &  23.05  & EK&  9.32 & 9.06 & 12.20    \\
 0915-118 & 0.054 & -24.46 &2.44 &  26.26 &  24.46  & WP&  8.85 & 9.17 & 12.26    \\
 0940-304 & 0.038 & -23.83 &2.59 &  24.09 &  23.48  & EK&  9.50 & 8.87 & 12.11    \\
 1043-290 & 0.060 & -24.71 &2.36 &  24.66 &  23.96  & EK&  8.98 & 9.30 & 12.57    \\
 1107-372 & 0.010 & -24.61 &2.47 &  23.07 &  21.64  & EK&  8.96 & 9.27 & 12.07    \\
 1123-351 & 0.032 & -24.67 &2.65 &  24.56 &  23.50  & EK&  9.75 & 9.29 & 12.72    \\
 1258-321 & 0.017 & -24.48 &2.42 &  24.00 &  23.10  & EK&  8.75 & 9.20 & 12.12    \\
 1333-337 & 0.013 & -24.55 &2.46 &  24.67 &  23.34  & WP&  9.04 & 9.24 & 12.17    \\
 1400-337 & 0.014 & -25.02 &2.49 &  23.52 &  22.11  & EK&  9.05 & 9.47 & 12.42    \\
 1404-267 & 0.022 & -24.00 &2.47 &  23.93 &  23.78  & EK&  8.98 & 8.96 & 12.19    \\
 1514+072 & 0.034 & -24.86 &2.43 &  24.70 &  24.31  & WP&  8.57 & 9.38 & 12.36    \\
 1521-300 & 0.020 & -22.01 &2.22 &  23.83 &  23.43  & EK&  7.82 & 7.97 & 11.04    \\
 2236-176 & 0.070 & -24.88 &2.39 &  25.08 &  23.35  & EK&  8.58 & 9.37 & 12.32    \\
 2333-327 & 0.052 & -23.77 &2.43 &  24.20 &  23.17  & EK&  8.77 & 8.83 & 11.90    \\
\hline
\multicolumn{9}{l}{Smith et al.(1990)} \\ 
\hline
  3C29 &  0.044 & -24.18 & 2.32 & 25.30 & 23.92   & WP  &  8.27 & 9.04 & 11.88    \\
  3C31 &  0.016 & -23.69 & 2.40 & 24.41 &   -     & KK  &  8.62 & 8.81 & 12.04    \\
  3C33 &  0.058 & -23.47 & 2.36 & 25.91 &   -     & KK  &  8.48 & 8.68 & 11.86    \\
  3C62 &  0.146 & -23.82 & 2.44 & 26.29 & 24.42   & WP  &  8.81 & 8.81 & 12.20    \\
3C76.1 &  0.032 & -23.21 & 2.30 & 24.82 &   -     & KK  &  8.20 & 8.56 & 11.45    \\
  3C78 &  0.028 & -24.60 & 2.42 & 25.13 & 24.55   & WP  &  8.73 & 9.26 & 11.99    \\
  3C84 &  0.017 & -24.37 & 2.40 & 25.81 &   -     & BB  &  8.61 & 9.15 & 12.08    \\
  3C88 &  0.030 & -23.49 & 2.28 & 24.91 & 23.81   & WP  &  8.08 & 8.70 & 11.99    \\
  3C89 &  0.138 & -24.10 & 2.40 & 25.87 &   -     & G1  &  8.64 & 8.95 & 12.47   \\
  3C98 &  0.030 & -22.83 & 2.24 & 25.32 &   -     & KK  &  7.91 & 8.37 & 11.56   \\
 3C120 &  0.033 & -23.78 & 2.30 & 25.63 & 25.24   & WP  &  8.20 & 8.84 & 11.81   \\
 3C192 &  0.061 & -23.21 & 2.28 & 25.54 &   -     & KK  &  8.12 & 8.55 & 11.57   \\
3C196.1 &  0.199 & -24.66 & 2.32 & 26.02 &   -    & G1  &  8.29 & 9.21 & 12.17   \\
 3C223 &  0.138 & -23.49 & 2.31 & 26.07 &   -     & KK  &  8.22 & 8.65 & 11.87   \\
  3C293 &0.046 &-23.95 & 2.27 &25.24 &   -        & KK  &  8.04 & 8.92 & 11.48   \\
  3C305 &0.042 &-24.16 & 2.25 &24.91 &   -        & BB  &  7.97 & 9.03 & 11.31   \\
\hline	     				    
\end{tabular}				    
\end{table*}

\addtocounter{table}{-1}
\begin{table*}
      \caption[]{The Sample A of radio galaxies (continue)}
\begin{tabular}{llcccccccc}
\hline
{\rm Object}&  z & M$_R$ & Log($\sigma$) &  Log($L_{48}$) &  Log($L_{48}$) & Ref.$^{a}$ & Log$\CMcal{M}_{BH}$ & Log$\CMcal{M}_{BH}$& Log$\CMcal {M}_{bulge}$ \\
& & host &  & total   & core  & &  from $\sigma$ & from M(bulge)  & \\
& & & km$s^{-1}$ &  (W/Hz) &  (W/Hz) & & $\CMcal{M}_{\odot}$ & $\CMcal{M}_{\odot}$  & $\CMcal{M}_{\odot}$ \\
(1)&(2) & (3) & (4) &(5) &(6) &(7) &(8) & (9) &(10) \\
\hline
      3C338 &0.031 & -24.9 & 2.46 &24.29 &   -     & BB  &  8.93 & 9.40 & 12.55   \\
      3C388 &0.090 &-24.92 & 2.56 &25.83 &   -     & KK  &  9.39 & 9.39 & 12.83   \\
PKS0634-206 &0.056 &-24.26 & 2.29 &25.7  & 23.23   & WP  &  8.15 & 9.07 & 11.52   \\
PKS2322-122 &0.081 &-24.33 & 2.35 &25.11 &   -     & G2  &  8.42 & 9.10 & 12.24   \\
      3C444 &0.152 &-24.77 & 2.19 &26.4  & $<$23.37& WP  &  7.69 & 9.28 & 12.21   \\
      3C449 &0.016 &-22.86 & 2.35 &24.25 &   -     & KK  &  8.42 & 8.39 & 12.19   \\
\hline
\multicolumn{9}{l}{Gonzalez-Serrano \& Carballo (2000)} \\
\hline
  NGC507 &  0.015 &-23.94 &2.52 &   -     &    - &  - &  9.18 & 8.93 & 12.10  \\
  NGC703 &  0.015 &-23.98 &2.38 &   -     &    - &  - &  8.57 & 8.95 & 12.21  \\
  gin116 &  0.033 &-24.01 &2.45 &  23.76  &    - & BB &  8.90 & 8.96 & 12.00  \\
 NGC4869 &  0.023 &-23.03 &2.30 &  23.38  &    - & BB &  8.19 & 8.48 & 11.45  \\
 NGC4874 &  0.025 &-24.89 &2.42 &  23.33  &    - & BB &  8.76 & 9.40 & 12.40  \\
 NGC6086 &  0.032 &-24.04 &2.51 &   -     &    - &  - &  9.14 & 8.98 & 12.26  \\
 NGC6137 &  0.031 &-25.57 & 2.47 &  23.87  &    - & BB &  8.97& 9.74 & 12.51  \\
\hline
\multicolumn{9}{l}{Ledlow \& Owen (1995)} \\
\hline
 0039-095B &  0.055 &-22.69 & 2.45 &  23.81  &  -    & G1 &  8.86 & 8.29 & 11.32  \\
  0053-015 &  0.038 &-24.03 & 2.47 &  24.66  &  -    & G1 &  8.98 & 8.97 & 12.20  \\
  0053-016 &  0.043 &-23.33 & 2.40 &  24.57  &  -    & G1 &  8.63 & 8.62 & 11.72  \\
  0110+152 &  0.044 &-23.87 & 2.29 &  24.36  &  -    & BB &  8.16 & 8.88 & 11.90  \\
  0112-000 &  0.045 &-22.94 & 2.40 &  23.58  &  -    & BB &  8.65 & 8.42 & 11.67  \\
  0122+084 &  0.049 &-24.71 & 2.56 &    -    &  -    &  - &  9.39 & 9.30 & 12.69  \\
  0147+360 &  0.017 &-23.00 & 2.38 &    -    &  -    &  - &  8.58 & 8.46 & 11.43  \\
  0306-237 &  0.067 &-23.44 & 2.40 &    -    &  -    &  - &  8.63 & 8.66 & 11.67  \\
  0431-133 &  0.033 &-24.46 & 2.43 &  23.32  &  -    & G2 &  8.78 & 9.18 & 12.31  \\
  0431-134 &  0.035 &-22.75 & 2.35 &  24.05  &  -    & G2 &  8.40 & 8.33 & 11.47  \\
  1510+076 &  0.043 &-22.39 & 2.53 &    -    &  -    &  - &  9.22 & 8.15 & 11.57  \\
  1520+087 &  0.034 &-24.20 & 2.34 &    -    &  -    &  - &  8.38 & 9.05 & 12.25  \\
 1602+178A &  0.031 &-22.80 & 2.33 &    -    &  -    &  - &  8.32 & 8.36 & 11.57  \\
  1610+296 &  0.032 &-23.88 & 2.51 &    -    &  -    &  - &  9.14 & 8.90 & 11.90  \\
 2322+143a &  0.045 &-22.36 &  2.31 &  23.99  &  -   & BB &  8.24 & 8.13 & 11.24  \\
 2335+267  &  0.030 &-24.60 & 2.54 &  25.13  &  -    & BB &  9.28 & 9.26 & 12.49  \\
\hline
\multicolumn{9}{l}{Faber et al. (1989)} \\
\hline
  NGC315 &  0.016 &-25.13 & 2.49 &  24.15 &    -    & BB &  9.31 & 9.53 & 12.57  \\
  NGC741 &  0.018 &-24.73 & 2.45 &  23.65 &    -    & G1 &  8.86 & 9.33 & 12.37  \\
 NGC4839 &  0.026 &-23.82 & 2.39 &  22.94 &    -    & BB &  8.71 & 8.87 & 12.20  \\
 NGC7626 &  0.010 &-22.98 & 2.51 &  23.14 &    -    & G1 &  8.51 & 8.46 & 11.85  \\
    3C40 &  0.017 &-23.66 & 2.23 &  24.43 &  23.15  & WP &  7.89 & 8.79 & 11.17  \\
\hline	     				    
\end{tabular}				    
\begin{list}{}{}
\item[$^{\mathrm{a)}}$]KK-K\"uehr et al. 1981; G1-Griffith et al. 1995;
G2-Griffith et al. 1994; BB-Becker et al. 1991
\end{list}
\end{table*}

We have considered the following two samples of radio galaxies:
\begin{itemize}

\item
{\it Sample A}: this consists of 72 radio-galaxies at z$<$0.2 with
available values of the absolute magnitude M$_R$, the effective radius
$R_e$ and the velocity dispersion $\sigma$.  We used this data in a
previous work (\cite{bettoni}) to study the Fundamental Plane of
RGs. In this sample 22 objects were observed by us, 22 galaxies were
taken from Smith et al. (1990)(SHI90), 16 from Ledlow \& Owen (1995),
7 from Gonzales-Serrano \& Carballo (2000) and 5 from Faber et
al. (1989)(FA89).  In Table~\ref{rg_bh} we report the relevant data
for the radio-galaxies in this sample: in columns 1, 2 we give the
name and the redshift; in column 3 we list the absolute R (Cousins)
magnitude to which we have also applied a correction to set the host
galaxy luminosity to the present epoch assuming a passive stellar
evolution for massive ellipticals (Bressan et al. 1994). Columns 4, 5,
6 and 7 report the Log($\sigma$), the total and core radio luminosity
at 4800 MHz and the reference for these radio data.  The velocity
dispersion is normalized to a circular aperture of metric radius
1.19$h^{-1}$ kpc (\cite{JFK}). This normalization is practically
equivalent to that adopted by Ferrarese \& Merritt (2000), who
consider $\sigma$ within an aperture radius of $r_e$/8. The average
difference between the two normalizations for our objects is $\leq$5
km/sec.

\item
{\it Sample B}: this consists of 79 radiogalaxies at z$<$0.1 for which
we secured homogeneous photometric and structural parameters
(\cite{FFS};~\cite{fede1},b).  These radio galaxies are extracted
from two complete surveys of radio sources (\cite{WP}: WP,
\cite{ekers}: EK) according to the specifications given in Fasano et
al. (1996), Sample B has in common 22 objects with Sample A. In
Table~\ref{rg_bh2} we report the relevant data for this sample. In
columns 1, 2 we give the name and the redshift; in columns 3, 4, 5 and
6 we list the absolute R magnitude, the total and core radio
luminosity at 4800 MHz and the reference for the radio data.
\end{itemize}

\section{The $\CMcal{M}_{BH}-M_R$(host) and $\CMcal{M}_{BH}-Log (\sigma)$ 
relations for normal, nearby ellipticals}

Using the dataset of nearby inactive ellipticals described in Section
2.1 we have derived the best fit of the relations Log($\CMcal
M_{BH}$)-$M_R$(host) and Log($\CMcal{M}_{BH}$)-Log($\sigma$). In order
to obtain a relationship between $\CMcal M_{BH}$ and M$_R$ usable for
$H_0$=50 we need to apply a color correction (to convert the B
magnitude into the R band) and a term that takes into account the
consistency between the adopted distances of nearby galaxies (see
Table~\ref{ngc_bh}) and the chosen value of $H_0$.  The latter term
can be written as 5Log($H_0'$/$H_0$), where $H_0'$=74 km/sec/Mpc
(Tonry et al. 2001).  The color correction was derived from LEDA with
the exception of NGC 4291 and of the three galaxies for which the B
magnitude has been obtained from RC3.  For these objects a standard
color ($B-R=1.60$) was assumed (\cite{Fukugita}).

The data used to fit the two relations $\CMcal{M}_{BH}$-M$_R$(host)
and $\CMcal{M}_{BH}-Log (\sigma)$ are shown in
Figure~\ref{MBH-MB-sigma}. The best fit to the data was derived
following the procedure outlined in Fasano \& Vio (1988), which takes
into account the individual measurement errors in both coordinates and
allows us to estimate the residual scatter $s_r$.

\begin{figure}
   \includegraphics[width=9cm]{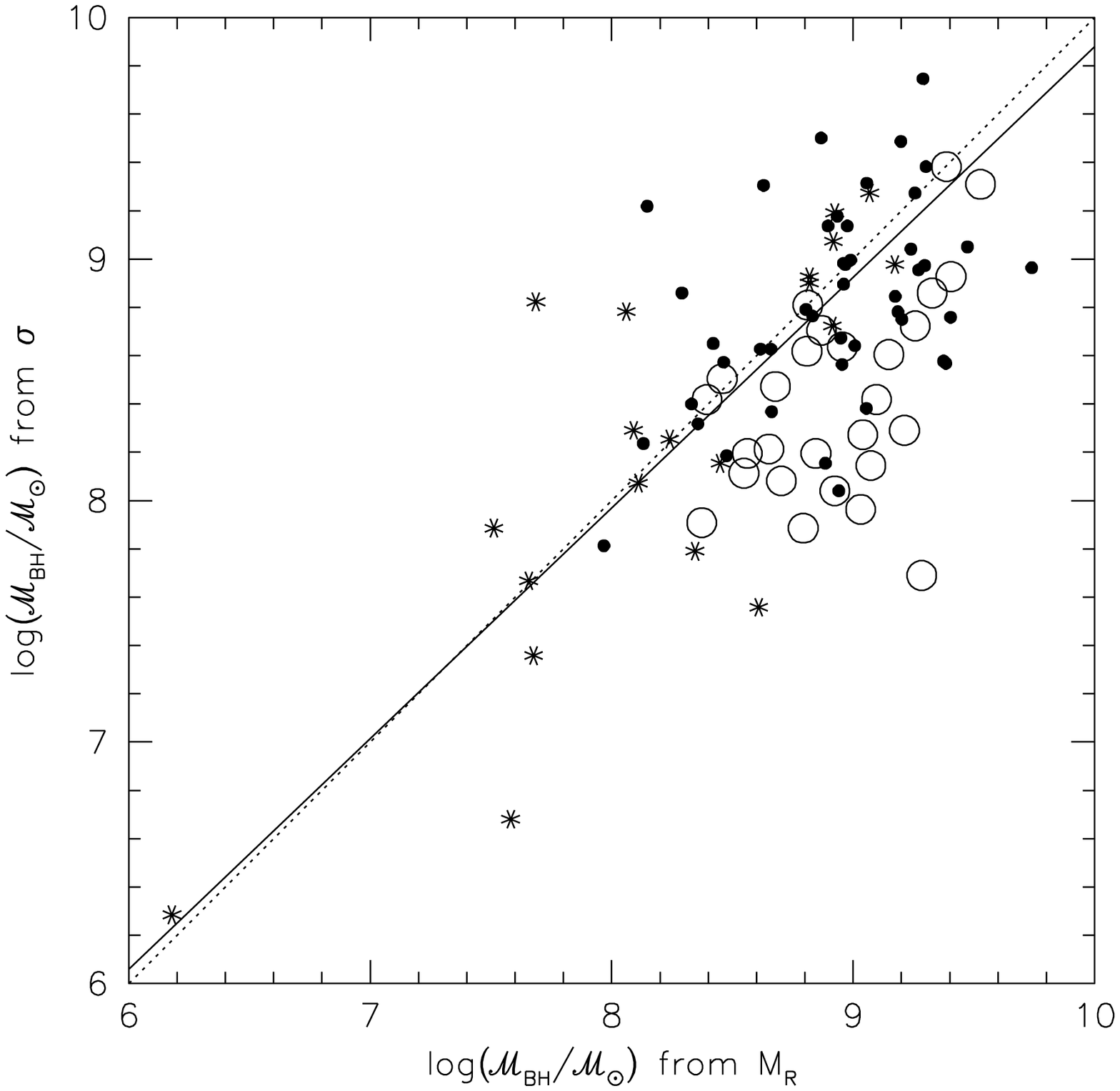}
\caption{The BH mass determinations obtained from $\CMcal{M}_{BH}-M_R$ 
and $\CMcal{M}_{BH}-Log (\sigma)$ for the Sample A of RG. A good
agreement is found for the reduced Sample A (filled circles), while the
objects from SHI90 and FA89 (big open circles) exhibit a systematic
deviation. For comparison we also plot $\CMcal M_{BH}$ for the sample
of 20 normal galaxies (asterisks). Solid line illustrates the best fit 
obtained for the reduced Sample A: $Log[\CMcal
M_{BH}(\sigma)]= 0.28 + 0.94\times Log[\CMcal M_{BH}(M_R)]$. The dotted
line is the one to one relation.}
\label{diff2}
\end{figure}
 
\begin{figure}
{\includegraphics[width=9cm]{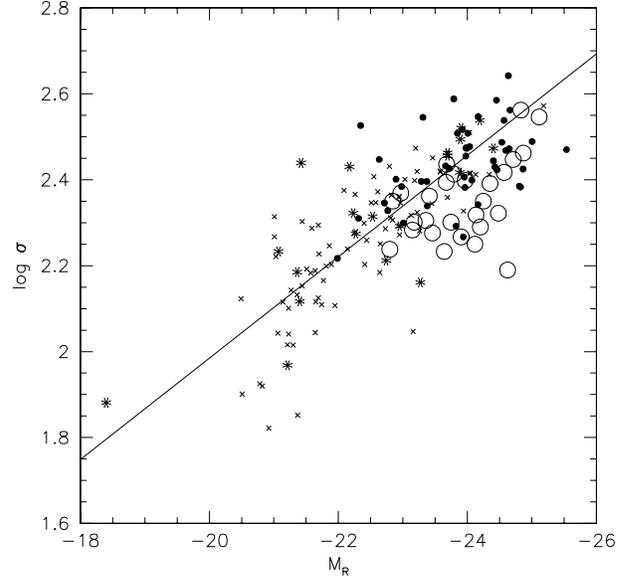}}
\caption{Faber-Jackson relation for RGs in Sample A and normal ellipticals.
RGs from SHI90 and FA89 (open circles) have systematically lower
$\sigma$ than the other RGs in the sample (filled circles). For comparison
we also plot the 20 inactive Es (asterisks) and data for early-type
galaxies from JFK96 (crosses). The solid line is our fit Log$\sigma$=-0.375
-0.118$\times$M$_R$ to the data of reduced Sample A and normal ellipticals
in the JFK96 samples.}
\label{fj}
\end{figure}

The two fitted relations are:

\begin{equation}
Log(\CMcal{M}_{BH}/\CMcal{M}_{\odot})=-0.50(\pm 0.06)\times M_{R}-3.00(\pm 1.35)~
\label{eq1}
\end{equation}
\vspace*{-0.3cm}
r.m.s.=0.39

\begin{equation}
Log(\CMcal{M}_{BH}/\CMcal{M}_\odot) = 4.55(\pm 0.49)\times Log~(\sigma) -2.27(\pm 1.13)~ 
\label{eq2}
\end{equation}
\vspace*{-0.3cm}
r.m.s.=0.41

\vspace*{0.4cm}
\noindent
where $M_{R}=M_{B}-(B-R)-5Log(H_0'/H_0)$. The residual scatters
of the two relations are 0.14($\pm$0.05) and 0.35($\pm$0.22), respectively.
In both cases they turn out to be consistent with
zero (within $3\sigma$), implying that the observed scatter is almost
entirely accounted for by the uncertainty of the measurements.  Both
relations are in good agreement with the last determinations by McLure
and Dunlop (2002) and Ferrarese (2002), respectively, once the different
cosmology is taken into account. However, at
variance with previous findings (\cite{FM},~\cite{kormendy2}),
luminosity and central velocity dispersion appear to be almost equally
correlated (very similar scatter) with the black-hole mass.  A similar
result was found by McLure \& Dunlop (2002) using a slightly different
sample of E-type objects. We note, however, that our scatter (0.39 dex)
is larger than that (0.31 dex) found by McLure \& Dunlop
(2002).  Since the data used by these authors are not published it is
not possible to further investigate the origin of this difference.


\section{Results}

\subsection{The BH~mass of radio~galaxies}

Here we use the relationships Log$\CMcal{M}_{BH}-M_R$(bulge) and
Log$\CMcal{M}_{BH}-Log (\sigma)$, derived in Section 3, (eq.~\ref{eq1}
and eq.~\ref{eq2}) to estimate the BH mass of the two samples (A and
B) of radio galaxies defined in section 2.2.  The BH masses derived
from $\sigma$ and from the M$_R$ are reported in columns 8 and 9 of
Table~\ref{rg_bh}. The comparison of the two
$\CMcal{M}_{BH}$ determinations for the Sample A is shown in
Figure~\ref{diff2}. We find that the mean values of $Log(\CMcal M_{BH})$ 
(8.66$\pm$0.44 from eq.~\ref{eq1} and 8.91$\pm$0.37 from eq.~\ref{eq2})
are significantly different. This is mainly attributable to
the systematically lower values of $\sigma$ for the 26 radio galaxies in the
sub-samples of SHI90 and FA89 (see figure~\ref{diff2}).
To further emphasize this point we plot in Fig~\ref{fj} the values of $\sigma$
versus M$_R$ (Faber-Jackson relation) for the RGs in Sample A and
for normal ellipticals from J\o rgensen et al. (1996, JFK96) sample. 

\begin{figure}
\includegraphics[width=14cm]{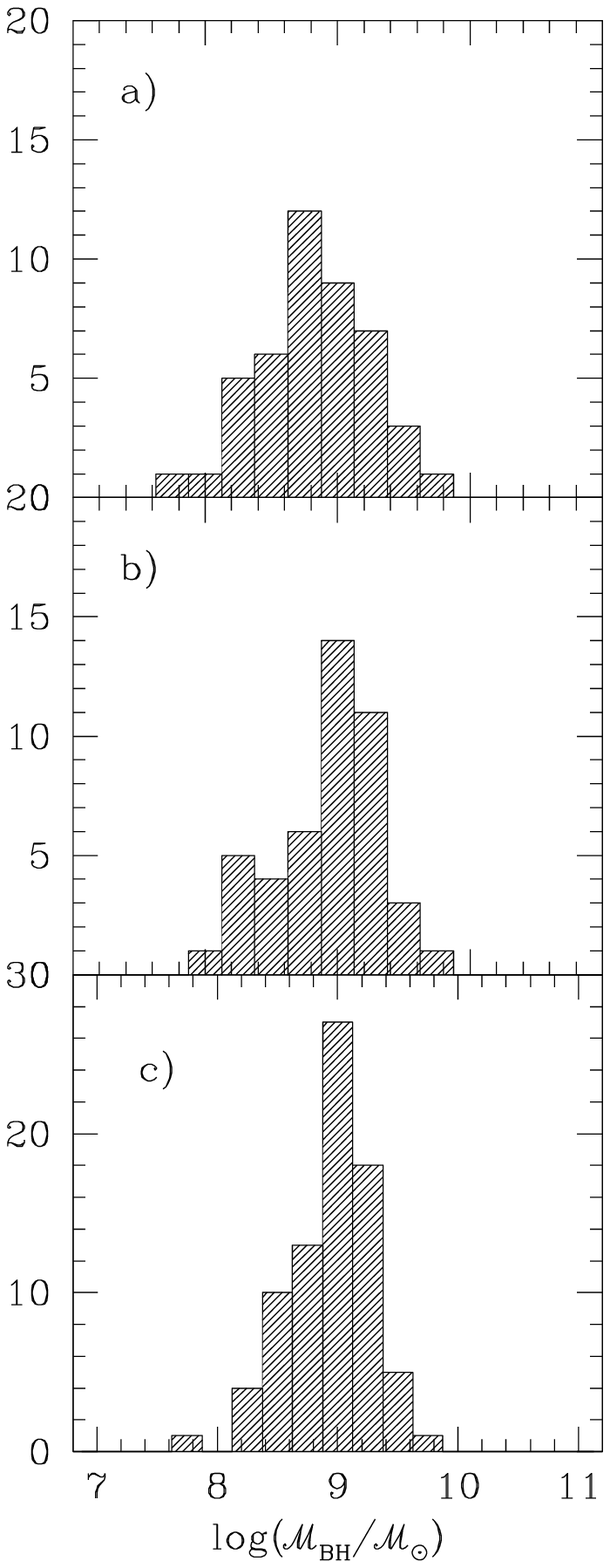}
\caption{a) The distribution of $\CMcal M_{BH}$ for 46 low z RGs
derived from measurements of $\sigma$(eq.~\ref{eq2}, reduces Sample A); b)The
distribution of $\CMcal M_{BH}$ for 46 low z RGs derived
from measurements of the bulge luminosity (eq.~\ref{eq1}, reduced Sample A); c)
The distribution of $\CMcal M_{BH}$ for 79 radiogalaxies derived from
the bulge luminosity (eq.~\ref{eq1}, Sample B)}
\label{histo}
\end{figure}

It is clearly apparent that RGs belonging to the SHI90 and FA89
samples deviate systematically from the overall relation derived from
fitting the data for the RGs in Sample A (excluding the SHI90 and FA89
galaxies) and for normal ellipticals from JFK96
(Log$\sigma$=-0.375-0.118$\times$M$_R$). Note also that the galaxies
in these datasets do not well agree with the other RGs of Sample A in
the Fundamental Plane (Bettoni et al. 2001). Therefore, in the
following analysis, we exclude the measurements of these 26 RGs.  If we
now consider the remaining 45 RGs in Sample~1 (hereafter reduced
Sample~1) we obtain: $<Log(\CMcal M_{BH})>_{\sigma}$=8.81$\pm$0.41 and
$<Log\CMcal M_{BH}>_{M_R}$=8.91$\pm$0.40.  For Sample~2 $\CMcal
M_{BH}$ was derived only from M$_R$ (eq~\ref{eq1}). These values are
given in Table~\ref{rg_bh2}. The mean value of $\CMcal M_{BH}$ 
$<Log\CMcal M_{BH}>_{M_R}$=8.94$\pm$0.34 is in good agreement with those
obtained for the reduced Sample A. In Figure~\ref{histo} the
distributions of $\CMcal{M}_{BH}$ for both samples are compared.

\begin{table*}
      \caption[]{The Sample B of radio active galaxies} 
\label{rg_bh2}
\begin{tabular}{llccccc}
\hline
{\rm Object}&  z  & M$_R$  &   Log($L_{48}$) &  Log($L_{48}$) & Ref. & log$\CMcal{M}_{BH}$ \\
& & host &  total &  core & & from M$_(bulge)$ \\
& & &  (W/Hz) &  (W/Hz) & & $\CMcal{M}_{\odot}$ \\
(1)&(2) & (3) & (4) &(5) &(6) &(7)  \\
\hline 			            
0005-199 &  0.121 & -24.46  &    25.25    &   24.00     & EK  &  9.20   \\
0013-316 &  0.107 & -24.51  &    24.85    &   23.44     & EK  &  9.23   \\
0023-333 &  0.05  & -24.57  &    24.67    &   23.05     & EK  &  9.26   \\
0034-014 &  0.073 & -23.67  &    25.60    &   24.87     & WP  &  8.81   \\
0055-016 &  0.045 & -24.10  &    25.30    &   23.93     & WP  &  9.02   \\
0123-016 &  0.018 & -24.02  &    24.43    &   23.15     & WP  &  8.98   \\
0131-367 &  0.03  & -24.13  &    25.21    &   23.18     & WP  &  9.04   \\
0229-208 &  0.089 & -24.10  &    25.25    &   24.54     & EK  &  9.02   \\
0247-207 &  0.087 & -25.21  &    25.03    &   23.91     & EK  &  9.58   \\
0255+058 &  0.023 & -22.95  &    24.66    &   22.96     & WP  &  8.45   \\
0257-398 &  0.066 & -23.67  &    24.94    &  $<$22.9    & EK  &  8.81    \\
0307-305 &  0.066 & -23.24  &    24.89    &   22.78     & EK  &  8.59   \\
0312-343 &  0.067 & -23.99  &    24.66    &   23.71     & EK  &  8.97   \\
0325+023 &  0.03  & -23.61  &    24.90    &   23.8      & WP  &  8.78   \\
0332-391 &  0.063 & -23.82  &    25.13    &   23.37     & EK  &  8.88   \\
0344-345 &  0.053 & -22.79  &    25.25    &   23.71     & EK  &  8.37 \\
0349-278 &  0.066 & -22.94  &    25.65    &   23.51     & EK  &  8.44 \\
0427-539 &  0.038 & -24.13  &    25.34    &   23.57     & WP  &  9.04 \\
0430+052 &  0.033 & -23.11  &    25.62    &   25.22     & WP  &  8.53 \\
0434-225 &  0.069 & -24.80  &    24.93    &   23.30     & EK  &  9.37 \\
0446-206 &  0.073 & -23.47  &    24.73    &   23.35     & EK  &  8.71 \\
0449-175 &  0.031 & -23.88  &    24.12    &   22.63     & EK  &  8.91 \\
0452-190 &  0.039 & -23.88  &    24.11    &   23.23     & EK  &  8.91 \\
0453-206 &  0.035 & -24.07  &    25.00    &   23.34     & WP  &  9.01 \\
0511-305 &  0.058 & -23.10  &    25.20    &   23.19     & EK  &  8.52 \\
0533-377 &  0.096 & -24.32  &    24.84    &   23.75     & EK  &  9.13 \\
0546-329 &  0.037 & -24.42  &    24.13    &   23.29     & EK  &  9.18 \\
0548-317 &  0.034 & -23.12  &    24.35    &   $<$22.49  & EK  &  8.53   \\
0620-526 &  0.051 & -24.92  &    25.17    &   24.49     & WP  &  9.43 \\
0625-354 &  0.055 & -24.43  &    25.46    &   24.92     & WP  &  9.19 \\
0625-536 &  0.054 & -25.20  &    25.39    &   23.75     & WP  &  9.57 \\
0634-205 &  0.056 & -23.78  &    25.70    &   23.23     & EK  &  8.86 \\
0712-349 &  0.044 & -24.16  &    24.17    &   23.34     & EK  &  9.05 \\
0718-340 &  0.029 & -24.07  &    24.55    &   23.05     & EK  &  9.01 \\
0806-103 &  0.11  & -23.96  &    25.98    &   24.50     & WP  &  8.95 \\
0915-118 &  0.054 & -24.36  &    26.26    &   24.46     & WP  &  9.15 \\
0940-304 &  0.038 & -23.70  &    24.09    &   23.48     & EK  &  8.82 \\
0945+076 &  0.086 & -23.11  &    25.96    &   24.05     & WP  &  8.53 \\
1002-320 &  0.089 & -24.20  &    24.93    &   $<$23.27  & EK  &  9.07   \\
\hline
\end{tabular}
\end{table*}

\addtocounter{table}{-1}
\begin{table*}
      \caption[]{The Sample B of radio active galaxies (continue)}
\begin{tabular}{llcccccc}
\hline
{\rm Object}&  z  & $M_R$ &   Log($L_{48}$) &  Log($L_{48}$) & Ref. & log$\CMcal{M}_{BH}$\\
& & host &  total &  core & &  M$_(bulge)$ \\
& & &  (W/Hz) & (W/Hz) & & $\CMcal{M}_{\odot}$\\
(1)&(2) & (3) & (4) &(5) &(6) &(7)  \\
\hline 			      
1043-290 &  0.06  & -24.62  &    24.66     &   23.96     & EK  & 9.28  \\
1053-282 &  0.061 & -23.99  &    25.18     &   24.29     & EK  & 8.97  \\
1056-360 &  0.07  & -23.54  &   25.06      &   24.05      &EK  & 8.74\\ 
1107-372 &  0.01  & -24.37  &   23.07      &   21.64      &EK  & 9.16\\ 
1123-351 &  0.032 & -24.53  &   24.56      &   23.50      &EK  & 9.23\\ 
1251-122 &  0.015 & -24.16  &   24.39      &   22.94      &WP  & 9.05\\ 
1251-289 &  0.057 & -25.31  &   24.55      &   $<$23.17   &EK  & 9.63   \\
1257-253 &  0.065 & -24.09  &   24.72      &   23.49      &EK  & 9.02 \\
1258-321 &  0.017 & -24.35  &   24.00      &   23.10      &EK  & 9.15 \\
1318-434 &  0.011 & -24.13  &   23.96      &   23.49      &WP  & 9.04 \\
1323-271 &  0.044 & -24.00  &   24.64      &   23.12      &EK  & 8.97 \\
1333-337 &  0.013 & -24.50  &   24.67      &   23.34      &WP  & 9.22 \\
1344-241 &  0.02  & -23.05  &   23.55      &   $<$21.85   &EK  & 8.50   \\
1354-251 &  0.038 & -23.46  &   24.29      &   22.51      &EK  & 8.70 \\
1400-337 &  0.014 & -24.90  &   23.52      &   22.11      &EK  & 9.42 \\
1404-267 &  0.022 & -23.91  &   23.93      &   23.78      &EK  & 8.93 \\
1514+072 &  0.034 & -24.77  &   24.70      &   24.31      &WP  & 9.36 \\
1521-300 &  0.02  & -21.78  &   23.83      &   23.43      &EK  & 7.86 \\
1637-771 &  0.041 & -23.41  &   25.29      &   24.14      &WP  & 8.68 \\
1717-009 &  0.031 & -22.49  &   25.94      &   23.70      &WP  & 8.22 \\
1733-565 &  0.098 & -23.68  &   26.19      &   25.49      &WP  & 8.81 \\
1928-340 &  0.098 & -24.63  &   24.94      &   23.98      &EK  & 9.29 \\
1929-397 &  0.073 & -25.04  &   25.39      &   23.54      &EK  & 9.49 \\
1949+023 &  0.059 & -23.69  &   25.58      &   23.2       &WP  & 8.82 \\
1954-552 &  0.058 & -23.33  &   25.55      &   23.89      &WP  & 8.64 \\
2013-308 &  0.088 & -24.69  &   25.01      &   23.56      &EK  & 9.32 \\
2031-359 &  0.088 & -24.54  &   25.33      &   23.64      &EK  & 9.24 \\
2040-267 &  0.041 & -24.09  &   24.83      &   23.37      &EK  & 9.02 \\
2058-282 &  0.039 & -24.24  &   25.13      &   23.63      &WP  & 9.09 \\
2059-311 &  0.039 & -24.23  &   24.13      &   23.09      &EK  & 9.09 \\
2104-256 &  0.038 & -24.20  &   25.46      &   23.59      &WP  & 9.07 \\
2128-388 &  0.018 & -23.20  &   23.76      &   22.46      &EK  & 8.57 \\
2158-380 &  0.034 & -23.16  &   24.47      &   $<$22.4    &EK  & 8.55   \\
2209-255 &  0.063 & -24.23  &   24.67      &   23.95      &EK  & 9.08 \\
2221-023 &  0.057 & -22.36  &   25.53      &   24.10      &WP  & 8.15 \\
2225-308 &  0.056 & -22.69  &   24.69      &   23.63      &EK  & 8.32 \\
2236-176 &  0.07  & -24.70  &   25.08      &   23.35      &EK  & 9.32\\
2333-327 &  0.052 & -23.77  &   24.20      &   23.17      &EK  & 8.86\\
2350-375 &  0.116 & -24.03  &   25.13      &   $<$23.41   &EK  & 8.99  \\
2353-184 &  0.073 & -23.92  &   24.88      &   $<$23.39   &EK  & 8.93  \\
\hline
\end{tabular}
\end{table*}								
									
\subsection{The $\CMcal M_{BH}-\CMcal M_{bulge}$ relation}

McLure and Dunlop (2002) and Laor (2001) investigated the relation
$\CMcal{M}_{BH}\propto\CMcal{M}^\eta_{bulge}$ for Seyfert galaxies and
powerful QSO in order to test its linearity (which means $\eta$=1). In
both papers the above relation is derived from $L_{bulge}$ under the
assumption of a mass-to-light ratio having the form
$\CMcal{M}/L\propto L^\Gamma$ but they reached different
conclusions. Laor (2001) claimed that the relation
$\CMcal{M}_{BH}-\CMcal{M}_{bulge}$ is not linear, while this was not
confirmed by McLure and Dunlop (2002). However, the latter authors
showed also that Laor data may be consistent with the linearity if one
adopts $\Gamma$=0.31 (J\o rgensen et al. 1996) instead of
$\Gamma$=0.18 (used by Laor).

\begin{figure}
\includegraphics[width=10cm]{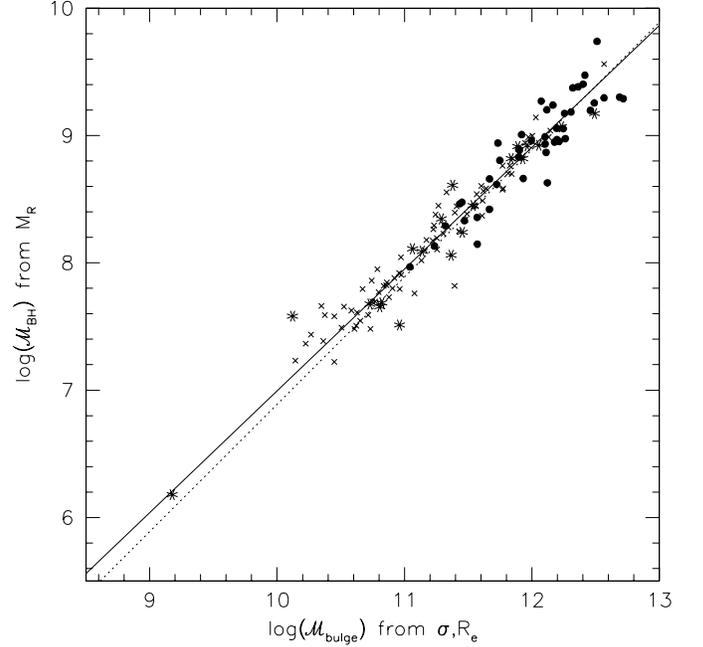}
\vspace{-1cm}
\caption{The relation between $\CMcal{M}_{BH}$ and $\CMcal{M}_{bulge}$ for 
RGs in the reduced Sample A (open circles), for the nearby inactive galaxies
with BH mass determination (asterisks) and
for JFK96 normal ellipticals (crosses). The solid line illustrates the
fit obtained for galaxies in the reduced Sample A and in JFK96 sample
(see eq.~\ref{eq3}). The dotted line represents the case of perfect
linearity ($\eta$=1) with $<Log(\CMcal{M}_{BH}/\CMcal{M}_{bulge})>=-3.11$.}
\label{fig5}
\end{figure}

Here we investigate the $\CMcal{M}_{BH}\propto\CMcal{M}^\eta_{bulge}$
relation for our samples of RGs.  We have computed the two involved
quantities ($\CMcal{M}_{BH}$ and $\CMcal M_{bulge}$) from $\sigma$,
M$_R$ and the effective radius of the galaxy ($R_e$) avoiding possible
spurious effects introduced in the analysis by the use of the same
variable to derive the two masses. We first derive $\CMcal{M}_{BH}$
from $M_R$ (eq.~\ref{eq1}) and $\CMcal M_{bulge}$ from $R_e$ and
$\sigma$ using the formula $\CMcal{M}_{bulge}$=$5G^{-1}\sigma^2 R_e$,
proposed by Bender et al. (1992).  These two determinations are
reported in columns 9 and 10 of Table~\ref{rg_bh}. Figure~\ref{fig5}
shows the linear fit to the $\CMcal{M}_{BH}$--$\CMcal{M}_{bulge}$
relation for objects in the reduced Sample A and for normal
ellipticals (JFK96), together with the data relative to the sample of
inactive galaxies for comparison. We find:

\begin{equation}
Log(\CMcal{M}_{BH})=0.96(\pm 0.03)\times Log(\CMcal{M}_{bulge})-2.56(\pm 0.34)~
\label{eq3}
\end{equation}

\noindent
This relation has $r.m.s.=0.14$, while the residual scatter (i.e.  the
scatter not accounted for by the errors) is $s_r$=0.03$\pm 0.012$,
which is consistent with zero. The slope ($\eta$=0.96) is practically
coincident with an almost perfect linearity ($\eta$=1; see the dotted
line in Figure~\ref{fig5}).  This suggests that a fundamental link is
present in the combined formation of BHs and spheroids that holds for
various types of active and inactive galaxies. We also note that,
given the coefficients of the fundamental plane (FP), the linearity of
the $\CMcal{M}_{BH}$-$\CMcal{M}_{bulge}$ relation turns out to be
directly linked with the particular value of the slope (A) of the
relation (Log$\CMcal{M}_{BH}$=A$\times$M$_R$ + B). In fact
the exponent $\eta$ of the $M_{BH}-M_{bulge}$ relation can be
written in the form: $\eta = -A\alpha /2\beta$, where $\alpha$ and
$\beta$ are the coefficients of the FP.  Using $\alpha$=1.242,
$\beta$=0.33 (Bettoni~et~al.~2001) and A=-0.5 it follows that,
$\eta$=0.94. Vice-versa, if perfect linearity is assumed using
the same $\alpha$, $\beta$ one obtains A=-0.53.

\begin{figure}
\resizebox{\hsize}{!}{\includegraphics{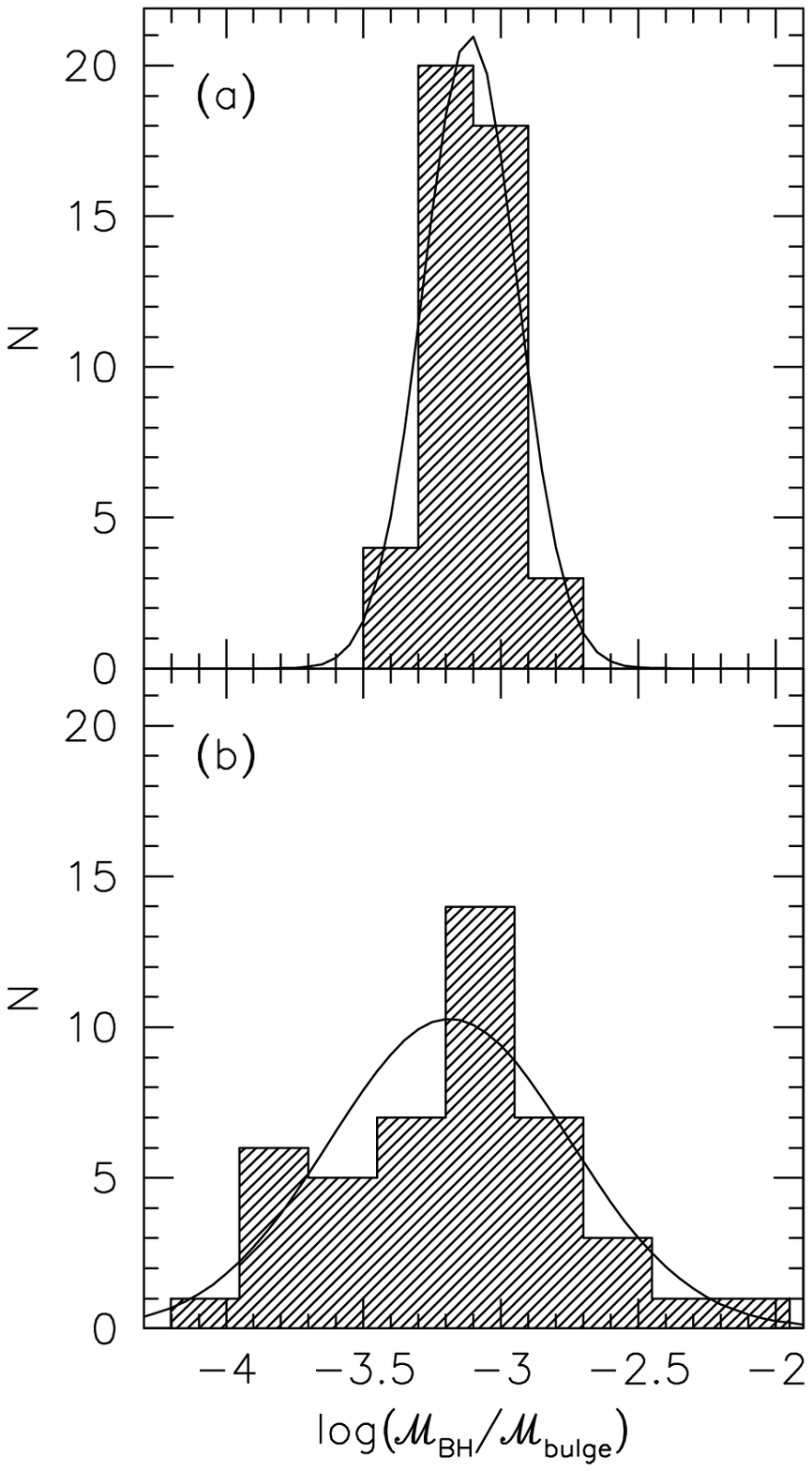}}
\caption{The distribution of the ratio $\CMcal{M}_{BH}/\CMcal{M}_{bulge}$ for 
the reduced Sample A of radiogalaxies: (a) $\CMcal{M}_{BH}$ from eq.~\ref{eq1}
and $\CMcal{M}_{bulge}$ from $\sigma$ and $R_e$; (b) 
$\CMcal{M}_{BH}$ from eq.~\ref{eq2} and $\CMcal{M}_{bulge}$ from
$L_{bulge}$ assuming $\CMcal{M}$/L$\propto L^{0.30}$. The
solid lines represent Gaussian distributions with the same average values and
$r.m.s.$ of the corresponding dataset (see text).}
\label{fig6}
\end{figure}

The distribution of the black-hole to bulge
mass ratio is shown in Figure~\ref{fig6} (panel a). The average value of
the mass ratio is $<Log(\CMcal{M}_{BH}/\CMcal{M}_{bulge})>=-3.11$, with 
$r.m.s.\sim 0.17$. The dispersion expected from just the
uncertainties on the measurements ($\sim 0.26$) is consistent
with that observed, suggesting that the intrinsic variance of the
ratio $\CMcal{M}_{BH}/\CMcal{M}_{bulge}$ is very small.

Alternatively we can derive the $\CMcal{M}_{BH}/\CMcal{M}_{bulge}$
ratio using $\CMcal{M}_{BH}$ derived from $\sigma$ and assuming
$\CMcal{M}_{bulge}$=0.0021$L^{1.30}$ (the coefficients are derived by
fitting the $\CMcal{M}$-L relation, in the R band,for our RG data
together with the ellipticals in JFK96). In this case we obtain
$<Log(\CMcal{M}_{BH}/\CMcal{M}_{bulge})>$=-3.16 with a dispersion of
0.43 (see Figure~\ref{fig6}b). These are in good agreement with those
derived for the inactive galaxy sample: -3.07 ($r.m.s.$=0.19) and
-3.05 ($r.m.s.$=0.50) for the two procedures respectively.  They are
also similar to previous determinations by Merritt and Ferrarese
(2001) who found -2.90$\pm$0.45 using the Magorrian et al. (1998)
galaxy sample and the $\CMcal{M}_{BH}$-$\sigma$ relation, and by
McLure and Dunlop (2002) that report -2.87$\pm$0.47 from their study
of the host galaxies of powerful quasars and use virial black-hole
masses.

\subsection{Relationship between $\CMcal M_{BH}$ and radio emission}

Based on a small number of nearby galaxies with known BH masses it was
suggested by Franceschini et al. (1998) that $\CMcal M_{BH}$ scales
with the total radio luminosity $L_{radio}$ at 5~GHz ($L_{radio}$
$\sim$ $\CMcal M_{BH}^{2.5}$). This is what would be expected by
Accretion-Dominated Accretion Flows (ADAF, see \cite{abram};
\cite{naran}) models.  This correlation appears to hold over
at least 3 order of magnitudes for $\CMcal M_{BH}$ and, given its
steepness, it was proposed as a tool to predict $\CMcal M_{BH}$ from
the simple observation of the radio flux.  Additional support for a
link between $L_{radio}$ and $\CMcal M_{BH}$ in various type of active
galaxies was lead by Laor (2001) and Lacy et al. (2001).

However, a recent analysis by Ho (2002) of this relationship, for
objects with a wide range of nuclear activity, argued that $\CMcal
M_{BH}$ is only loosely related with the radio power (both total and
core emission).  The poor correlation observed between the two
quantities could therefore arise from indirect relations between radio
luminosity, bulge mass and $\CMcal M_{BH}$ and has no practical power
to predict $\CMcal M_{BH}$ from the radio luminosity.

\begin{figure}
\includegraphics[width=10cm]{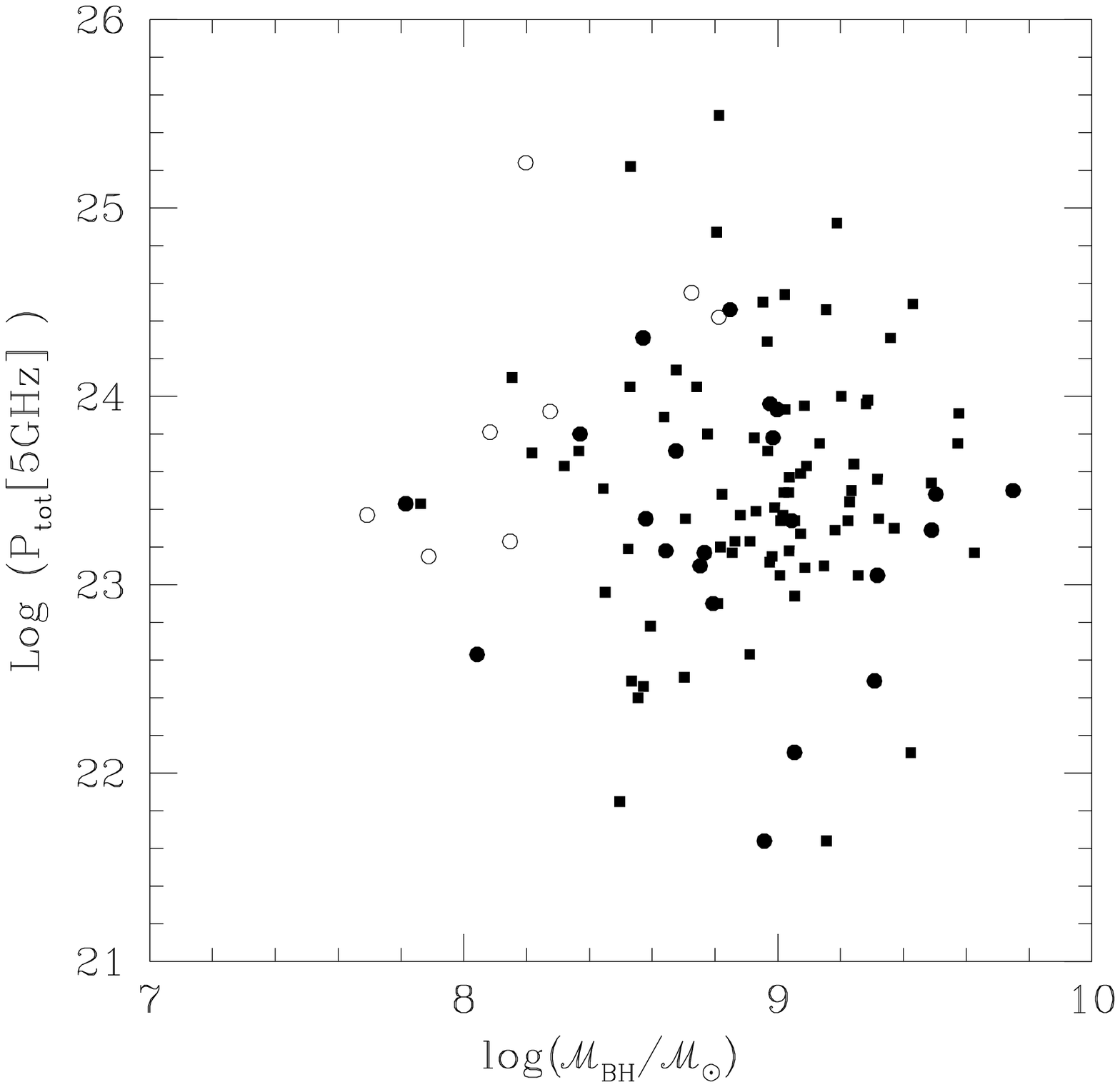}
\caption{$\CMcal M_{BH}$-Log$P_{tot}$(5 GHz) plane for RGs in 
Sample~1 (filled circles, reduced Sample A, open circles data from
SHI90 and FA89) and Sample~2 (filled squares). $\CMcal M_{BH}$ is
derived from $\sigma$ (see data in
Tables~\ref{rg_bh} and \ref{rg_bh2}). The two
quantities are clearly not correlated.}
\label{radio1}
\end{figure}

We have investigated this issue using our samples of radiogalaxies for
which radio power ($P_{tot}$) at 5 GHz and $\CMcal M_{BH}$ are available (see
Table~\ref{rg_bh}). In figure~\ref{radio1} we show 
the data for our radio galaxies in the plane
$\CMcal M_{BH}$-LogP$_{5GHz}$(total).  The two quantities are not
correlated: Spearman correlation coefficient -0.174, with significance
of nonzero correlation 0.038. The same result is found using the
core radio power. Similarly, no significant correlation is
found between the host galaxy mass $\CMcal M_{bulge}$ and the radio
power, in agreement with the fact that $\CMcal M_{BH}$ is tightly
related to $\CMcal M_{bulge}$.

\begin{figure}
\includegraphics[width=10cm]{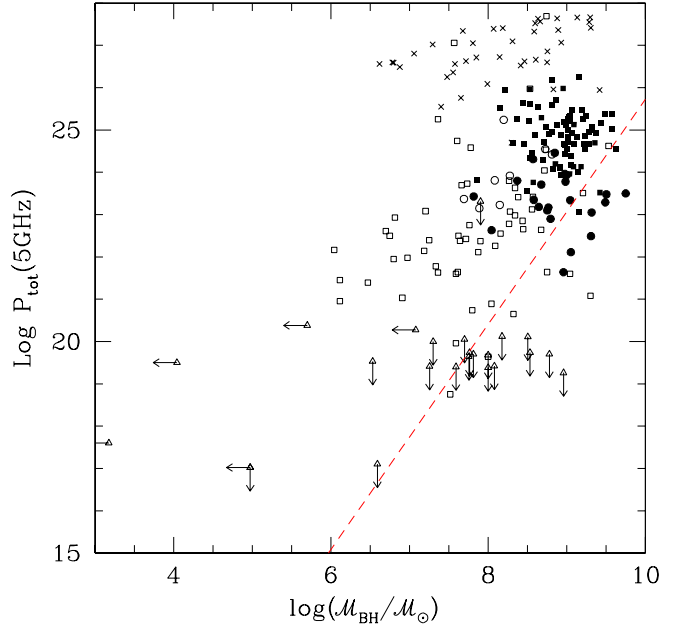}
\caption{$\CMcal M_{BH}$-LogP(total) relation for the samples A (filled 
circles, reduced Sample A, open circles data from
SHI90 and FA89) and B (filled squares) of radiogalaxies, compared with
the objects (opens squares and triangles) studied by Ho (2002) and
with radio loud quasars (crosses) investigated by Oshlack et
al. (2002). The dashed line is the Franceschini et al. 1996 relation
($P_{tot}\propto \CMcal M_{BH}^{2.5}$)}
\label{radio2}
\end{figure}

In Figure~\ref{radio2} we compare the distribution of our RGs over the
plane $\CMcal M_{BH}$ -- radio power with that of other samples of galaxies at
various levels of nuclear activity investigated by Ho (2002) and
Oshlack et al. (2002). Note that the latter data refer to flat
spectrum radio-sources whose flux is likely enhanced by Doppler
boosting (beamed sources). Their observed luminosity therefore tends
to occupy the higher region of the $P_{tot}$-$\CMcal M_{BH}$ plane
(see also Jarvis \& McLure 2002). Our data for RG are in
agreement with the analysis by Ho (2002) and enforce his conclusion
that radio power is poorly correlated with the mass of the central BH
and therefore that the latter can not practically be predicted by the
measurement of the radio flux.  If instead of the total radio
luminosity we use the radio power of the core this picture remains
basically unchanged. All together the points in Figure~\ref{radio2}
appear to follow the trend of increasing $\CMcal M_{BH}$ with higher
radio luminosity. This behavior is consistent with the suggestion by
Dunlop et al. (2002) that both active and inactive galaxies fall
between two envelopes (following the $P_{tot}\propto$ $\CMcal
M_{BH}^{2.5}$ law) in the $\CMcal M_{BH}$-$P_{tot}$ plane.  On the
other hand the scatter of the points is rather large and do not
produce a significant correlation of the two quantities. Our objects
cover the top-right part of the diagram where more massive BH and
powerful radio sources are expected.  Most of our points and those of
Ho (2002) lie above the relation proposed by Franceschini et
al. (1998) by 2-3 order of magnitude.  At low accretion rates ADAF
models foresee that radio power depends on $\CMcal M_{BH}$ as
$P_r\propto \CMcal M_{BH}^{2.5}$. The location of points well above
such relation may simply reflect the much higher level of activity
(and likely of accretion).

\section{Summary and Conclusions}
			 
We used of empirical relations between $\CMcal M_{BH}$ and either the
velocity dispersion ($\sigma$) or the bulge luminosity M$_R$(bulge)
derived for nearby early type galaxies to infer the mass of the
central BH of low redshift radio galaxies.

The main conclusions of this study are: 

1) Using only nearby galaxies of E-type morphology, for which BH
masses are available, the two relationships $\CMcal M_{BH}$ - $\sigma$
and $\CMcal M_{BH}$ - M$_{bulge}$ exhibit a similar scatter ($\sim$0.4 in
Log$\CMcal M_{BH}$) which is also consistent with that expected from
the estimated errors  on the involved parameters. 

2) We showed that when the dependence on the adopted cosmology (for
the $\CMcal M_{BH}$ - M$_{bulge}$ relation) is properly taken into
account the two relations predict the same value of the central BH
mass (within the expected uncertainties). This means that the BH mass
can be reliably estimated from the observable parameter M$_R$(bulge)
which is more easily measurable than $\sigma$.

3) The inferred BH mass of low redshift radio galaxies is in the range
5$\times10^7$ to 5$\times10^9$ $\CMcal {M}_{\odot}$. 

4) We found that the central BH mass is linearly correlated with
that of its host galaxy the average Black Hole to bulge
mass ratio is $<Log[\CMcal{M}_{BH}/\CMcal{M}_{bulge}]>\sim$ -3.1.

5) The total (or core) radio power at 5~GHz is not correlated neither
with the mass of the central BH nor with that of the galaxy; the radio
power is always in excess by 2-3 order of magnitude with respect to
what would be expected from low accretion rate models (ADAF).

\begin{acknowledgements}
We wish to acknowledge the anonymous referee for valuable comments which lead 
to the improvement of this paper. This work has received partial support
under contracts COIN2001/028773-004, ASI-IR-35 and ASI-IR-73-01. 
\end{acknowledgements}

\end{document}